\documentclass[aps,prl,twocolumn,showpacs,superscriptaddress,groupedaddress]{revtex4-1}  % for review and submission
\usepackage{graphicx}  % needed for figures
\usepackage{dcolumn}   % needed for some tables
\usepackage{bm}        % for math
\usepackage{amssymb}   % for math
\usepackage{atlasphysics}
\usepackage{hyperref}
\def\integlumi{42}
\def\explimitminus{5.1}
\def\explimitplus{4.8}
\def\obslimitminus{4.9}
\def\obslimitplus{4.5}
\def\ttbar{\ensuremath{t\bar{t}}}

\begin{document}

\widetext
CERN-PH-EP-2011-052; submitted to Phys. Rev. D Rapid Communications.

\title{Search for Contact Interactions in Dimuon Events from
 pp Collisions at $\sqrt{s} = 7$~\TeV{} with the ATLAS Detector
}

\author{The ATLAS Collaboration}

\begin{abstract}
A search for contact interactions has been performed using dimuon events
recorded with the ATLAS detector in proton-proton collisions
at $\sqrt{s}=7$~\TeV{}.
The data sample corresponds to an integrated luminosity of
$\integlumi$~pb$^{-1}$. No significant deviation from the Standard Model is observed in the
dimuon mass spectrum, allowing the following 95\% C.L. limits to be set on the energy
scale of contact interactions: 
$\Lambda > \obslimitminus$~\TeV{} ($\obslimitplus$~\TeV{})
for constructive (destructive) interference in the left-left
isoscalar compositeness model.
These limits are the most stringent to date for $\mu \mu q q$ contact interactions.
\end{abstract}

\maketitle

%\linenumbers

%\section{\label{sec:level1}First-level heading}
% sections are not used for PRL papers

% what is the physics?
Phenomena beyond the Standard Model (SM), such as large extra spatial dimensions
in the ADD model~\cite{ADD} or quark/lepton compositeness~\cite{ELP}, 
may be described as a four-fermion contact interaction (CI) in the low energy limit.
Such an approach is similar to that used by Fermi to describe nuclear $\beta$~decay~\cite{Fermi}
long before the discovery of the $W$ boson. One can describe
a new interaction at a higher energy scale with an effective
Lagrangian of the form~\cite{ELP}

\begin{equation}\label{lagrangian}
\begin{array}{r@{\,}c@{}c@{\,}l@{\,}l}
\mathcal L = \frac{g^2}{2\Lambda^2}\;[&& \eta_{LL}&\, \overline\psi_L\gamma_{\mu}\psi_L\,\overline\psi_L\gamma^{\mu}\psi_L \nonumber \\
& +&\eta_{RR}& \overline\psi_R\gamma_{\mu}\psi_R \,\overline\psi_R\gamma^{\mu}\psi_R \\
&+&2\eta_{LR}& \overline\psi_L\gamma_{\mu}\psi_L \,\overline\psi_R\gamma^{\mu}\psi_R&]\: ,\nonumber
\end{array}
\end{equation}
where $g$ is a coupling constant,
$\Lambda$ is the energy scale below which fermion constituents
are bound (in the context of compositeness models),
and $\psi_{L,R}$ are left-handed and right-handed fermion fields, respectively.
The scale $\Lambda$ is defined by the choices $g^2/4\pi = 1$
and $\eta_{LL},\eta_{LR},\eta_{RR} = \pm 1$.
Different choices of the parameters $\eta_{LL}$, $\eta_{LR}$ and $\eta_{RR}$
determine the helicity structure of the new interaction. 
For example,
the analysis presented in this Letter applies specifically
to the left-left isoscalar model (LLIM) commonly used as a benchmark
for contact interactions searches~\cite{PDG}.
This model is defined by setting 
$\eta_{LL}=\pm1$ and $\eta_{LR}=\eta_{RR}=0$.
With the introduction of a contact interaction, the differential cross section
for the process $q \bar{q} \to \mu^+\mu^-$ becomes
\begin{equation}\label{cross_section}
\frac{d\sigma}{dm_{\mu\mu}} = \frac{d\sigma_{DY}}{dm_{\mu\mu}}
 - \eta_{LL}\frac{F_I(m_{\mu\mu})}{\Lambda^2} + \frac{F_C(m_{\mu\mu})}{\Lambda^4}~,
\end{equation}
where $m_{\mu\mu}$ is the final-state dimuon mass.
The expression above includes a SM Drell-Yan (DY) term, 
as well as DY-CI interference ($F_I$) and pure contact interaction ($F_C$) terms.
DY here incorporates both photon and $Z^0$ boson contributions.

% what is this letter about?
This Letter presents the results of a search for contact interactions
in the dimuon channel, taking
advantage of the high $pp$ collision energy of the LHC
and the capabilities of ATLAS to detect and measure muons.
The search strategy focuses on identifying a deviation from the
SM in the dimuon mass spectrum, which is expected to be
dominated by DY.
Contributions from a new interaction would undergo either
constructive ($\eta_{LL} = -1$) or destructive ($\eta_{LL} = +1$) interference
with the DY contribution.
If present, a signal would result in a broad deviation from the SM
expectation rather than a peak in the mass spectrum.
Given current experimental bounds on $\Lambda$ (see below), such a deviation would appear 
at masses well above the $Z^0$ boson peak.
Therefore, the measurement requires excellent
muon identification and reconstruction at high momentum.
A separate Letter presents the results of a search for new heavy resonances in the
dimuon mass spectrum~\cite{Zprime}.
Previous searches for contact interactions have been carried out 
in neutrino scattering~\cite{NuTeV},
as well as at electron-positron~\cite{ALEPH,DELPHI,L3,OPAL}, 
electron-proton~\cite{H1,ZEUS} and 
hadron colliders~\cite{Abe97,Affolder01,Abulencia06,Abbott99,Abazov09,dijet_CMS_3ipb,dijet_ATLAS_3ipb,dijet_CMS_36ipb,dijet_ATLAS_36ipb}.
For the channel under study, the best limits in the LLIM are 
$\Lambda^- > 4.2$~\TeV{} for constructive interference
and $\Lambda^+ > 2.9$~\TeV{} for destructive interference,
at 95\% C.L.~\cite{Abe97}.

% ATLAS description
ATLAS is a multipurpose particle detector~\cite{detector} designed
for physics at the \TeV{} scale.
Charged particle tracking is provided by an inner detector consisting
of a pixel detector, a silicon-strip tracker and
a transition radiation tracker, immersed in a 2~T solenoidal magnetic field.
A high-granularity liquid-argon electromagnetic calorimeter
surrounds the solenoid.
Hadron calorimetry is provided by an iron-scintillator tile calorimeter
in the central rapidity range and a liquid-argon calorimeter in the
endcap and forward rapidity range.
A key detector component for this analysis is the muon spectrometer,
which is designed to identify muons and measure both their trajectories and
momenta with high accuracy: the design momentum resolution is 10\% 
at momenta transverse to the beam line ($p_T$) of 1~\TeV{}.
The muon spectrometer comprises three toroidal magnet systems
consisting of eight coils each with
a bending power $\int B {\mathrm d}\ell = 1-7.5$~Tm, a trigger system 
consisting of both resistive plate chambers and thin-gap chambers, and
a set of precision monitored drift tubes and cathode strip chambers 
with a single-hit spatial resolution better than 100~$\mu$m to
accurately measure muon curvature.
Precision chambers are continuously monitored by an optical alignment system
designed to determine relative chamber positions
to an accuracy of 50~$\mu$m or better.

The data sample for this analysis was collected during LHC operations in 2010
and corresponds to a total integrated luminosity of $\integlumi$~pb$^{-1}$
collected with stable beam conditions and fully operational
inner detector and muon spectrometer systems.
Events with muons were selected by requiring the presence of at least
one high-momentum muon passing all three rejection levels of the
muon trigger system. The $p_T$ threshold was initially set to 10~\GeV{}
but was raised to 13~\GeV{} in the later parts of the data taking due
to increasing luminosity.

% event selection
This analysis follows the same event selection as the search for new
heavy resonances. A summary is provided below; see Ref.~\cite{Zprime}
for a more complete description. 
Events with a good primary vertex are selected to suppress cosmic-ray
events.
Muon tracks reconstructed independently in the inner detector and
muon spectrometer are combined with a fit to all associated hits,
taking the energy loss in the calorimeter into account.
The energy loss estimate uses either the parameterized expected
energy loss or the energy measured in the calorimeter 
if this energy significantly exceeds the most probable energy loss.
The combined tracks are required to have hits in all inner detector
tracking systems, have at least one hit in the non-bending plane
and at least three hits in each of the
inner, middle and outer precision chambers of the muon spectrometer.
Tracks passing through poorly aligned chambers are rejected.
The above hit requirements guarantee a reliable momentum measurement and good modeling by
the detector simulation.
Muon tracks are required to have
$p_T > 25$~\GeV{},
pseudorapidity $|\eta| < 2.4$~\cite{coordinates} 
to be within the acceptance of the inner detector
tracking and muon spectrometer trigger systems,
and a relative track isolation 
$\sum p_T^i / p_T < 0.05$,
where the sum is over all inner detector tracks $i$ within
a $\Delta R = \sqrt{\Delta \eta^2 + \Delta \phi^2}$ cone of 0.3
around the muon trajectory,
to suppress backgrounds from heavy flavor decays.
Additional requirements are placed on the impact parameter of the muon
track to reduce cosmic-ray backgrounds to a negligible level.
Finally, dimuon candidates are formed from all pairs of opposite-charge muons satisfying
the above criteria and the mass of those pairs is required to be greater than 70~\GeV{}.
There are 7743 dimuon events passing all selection requirements.

% MC simulation
Drell-Yan, $W$+jets and multi-jet events were generated
with {\sc pythia} 6.421~\cite{Pythia} and 
{\sc mrst2007} LO$^\ast$ parton distribution functions (PDFs)~\cite{MRST}.
Diboson ($WW$, $WZ$, and $ZZ$) events were produced with
{\sc herwig} 6.510~\cite{Herwig} and {\sc mrst2007} LO$^\ast$ PDFs.
In the case of $\ttbar$, events were
generated with {\sc mc@nlo} 3.41~\cite{MCatNLO} to compute matrix elements,
{\sc jimmy} 4.31~\cite{Jimmy} to simulate the underlying event,
{\sc herwig} 6.510 to model parton showering and hadronization,
and {\sc cteq} 6.6~\cite{CTEQ} for PDFs.
For signal samples, {\sc pythia} 6.421 was used to produce DY and CI
simultaneously in order to properly account for the interference between the two processes.
A mass-dependent QCD K-factor corresponding to the ratio between NNLO~\cite{Catani09}
and {\sc pythia} LO$^\ast$ DY differential cross sections was applied to these signal samples
as well as pure DY samples.
Similarly, a mass-dependent electroweak K-factor was applied to account for
higher order electroweak effects due to virtual gauge boson loops~\cite{Horace}.
This correction was only applied to the DY cross section since the new physics
included in the CI term has unknown couplings to SM gauge bosons.
The QCD (electroweak) K-factor varies between 1.16 (1.04) at low dimuon mass
and 0.86 (0.85) at a mass of 2~\TeV{}.
The response of the ATLAS detector to these generated event samples
was simulated with {\sc geant} 4~\cite{ATLASsimul,Geant}.

\begin{figure}
  \includegraphics[width=0.48\textwidth]{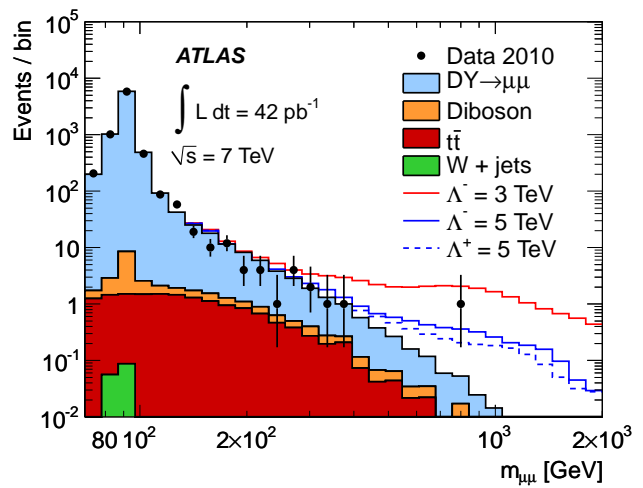}
  \caption{\label{fig:dimuon_mass} Dimuon invariant mass distribution for data (points)
  and Monte Carlo (histograms). The red (blue) line corresponds to the distribution
  expected in the presence of contact interactions with $\Lambda^- =$ 3~\TeV{} (5~\TeV{})
  for constructive interference. The dashed blue line corresponds to $\Lambda^+ =$ 5~\TeV{}
  for destructive interference.}
\end{figure}

% results
Figure~\ref{fig:dimuon_mass} shows the dimuon mass distribution for all selected events
along with the predicted contributions from SM processes and
CI for selected $\Lambda$ values.
Predictions for the various background processes are extracted from the 
Monte Carlo (MC) simulation. 
Besides the dominant DY contribution, we also account for a small
dimuon yield from $\ttbar$ and diboson production.
The small predicted yield from $\ttbar$ has been confirmed in the data
by selecting events with high-mass electron-muon pairs, see Ref.~\cite{Zprime}.
Backgrounds from W production are effectively suppressed by requiring 
two selected muons in the event.
Likewise, multi-jet backgrounds are reduced to a negligible amount
($< 0.1$ events in the selected sample) 
by the muon $p_T$ and isolation requirements.

Table~\ref{tab:dimuon_yields} presents the number of events in different
bins of dimuon mass for data and MC simulation.
The sum of MC predictions is normalized to the number of data events
in the $Z^0$ peak mass region between 70 and 110~\GeV{}.
It should be noted that, prior to normalization, data and MC event yields 
agree within the uncertainty in the integrated luminosity.
This normalization procedure removes sensitivity
to mass-independent uncertainties such as the luminosity uncertainty.
The overall acceptance of the selection is estimated to be 36\% 
for simulated DY events in the signal region defined by $m_{\mu\mu} > 150$~\GeV{}.
% Background_Table.tex
\begin{table*}[t]
\newcommand\T{\rule{0pt}{2.6ex}}
\newcommand\B{\rule[-1.2ex]{0pt}{0pt}}
\caption{Expected and observed number of events in the dimuon channel.
The errors quoted originate from the limited MC statistics.
Entries of 0.0 indicate a value $< 0.05$.
}
\label{tab:dimuon_yields}
\centering
%\tiny
\scriptsize
%\small\addtolength{\tabcolsep}{-5pt}
%\resizebox{\columnwidth}{!}{
\begin{tabular}{cr@{\ $\pm$\ }l r@{\ $\pm$\ }l r@{\ $\pm$\ }l r@{\ $\pm$\ }l r@{\ $\pm$\ }l r@{\ $\pm$\ }l r@{\ $\pm$\ }l ccccc}
\hline
\hline
$m_{\mu\mu}$ [\GeV{}] \T \B & \multicolumn{2}{c}{70-110} & \multicolumn{2}{c}{110-130} & \multicolumn{2}{c}{130-150} & \multicolumn{2}{c}{150-170} & \multicolumn{2}{c}{170-200} & \multicolumn{2}{c}{200-240} & \multicolumn{2}{c}{240-300} & 300-400 & 400-550 & 550-800 & 800-1200 & 1200-2000 \\
\hline
DY \T  & 7547 & 7 & 98.4 & 0.8 & 33.4 & 0.5 & 17.2 & 0.3 & 12.8 & 0.3 & 7.8 & 0.2 &  5.1 & 0.1 & 2.5 $\pm$ 0.0 & 1.0 $\pm$ 0.0 & 0.3 $\pm$ 0.0 & 0.1 $\pm$ 0.0 & 0.0 $\pm$ 0.0 \\
$\ttbar$ \T & 6.0 & 0.2 & 2.4 & 0.1 & 1.7 & 0.1 & 1.2 & 0.0 & 1.2 & 0.0 & 1.0 & 0.0 & 0.73 & 0.0 & 0.4 $\pm$ 0.0 & 0.1 $\pm$ 0.0 & 0.0 $\pm$ 0.0 & 0.0 $\pm$ 0.0 & 0.0 $\pm$ 0.0 \\
Diboson \T & 10.1 & 0.1 & 0.8 & 0.2 & 0.6 & 0.0 & 0.5 & 0.0 & 0.4 & 0.0 & 0.3 & 0.0 & 0.24 & 0.0 & 0.2 $\pm$ 0.0 & 0.1 $\pm$ 0.0 & 0.0 $\pm$ 0.0 & 0.0 $\pm$ 0.0 & 0.0 $\pm$ 0.0 \\
$W$+jets \T \B & 0.14 & 0.08 & 0.0 & 0.0 & 0.0 & 0.0 & 0.0 & 0.0 & 0.0 & 0.0 & 0.0 & 0.0 & 0.0 & 0.0 & 0.0 $\pm$ 0.0 & 0.0 $\pm$ 0.0 & 0.0 $\pm$ 0.0 & 0.0 $\pm$ 0.0 & 0.0 $\pm$ 0.0 \\
%QCD \T &  0.05 $\pm$ 0.03  &  0.01 $\pm$ 0.01  &  0.01 $\pm$ 0.01  &  0.01 $\pm$ 0.01  &  0.00 $\pm$ 0.00  &  0.00 $\pm$ 0.00  &  0.00 $\pm$ 0.00  &  0.00 $\pm$ 0.00  &  0.00 $\pm$ 0.00  &  0.00 $\pm$ 0.00  &  0.00 $\pm$ 0.00  &  0.00 $\pm$ 0.00 \\
\hline
Total \T \B & 7563 & 7  &  101.6 & 0.8  &  35.7 & 0.5  &  18.9 & 0.3  &  14.4 & 0.3  &  9.1 & 0.2 &  6.0 & 0.1  &  3.0 $\pm$ 0.1  &  1.2 $\pm$ 0.0  &  0.3 $\pm$ 0.0  &  0.1 $\pm$ 0.0  &  0.0 $\pm$ 0.0  \\
\hline
Data \T \B & \multicolumn{2}{c}{7563}  &  \multicolumn{2}{c}{101}  &  \multicolumn{2}{c}{41}  &  \multicolumn{2}{c}{11}  &  \multicolumn{2}{c}{11}  &  \multicolumn{2}{c}{7}  &  \multicolumn{2}{c}{6}  &  2  &  0  &  1  &  0  &  0 \\
\hline
\hline
\end{tabular}
%}
\end{table*}

\begin{table*}[ht]
\newcommand\T{\rule{0pt}{2.6ex}}
\newcommand\B{\rule[-1.2ex]{0pt}{0pt}}
\caption{Expected number of events in the signal region of the analysis
  for various contact interaction scales with constructive ($\Lambda^-$)
  and destructive ($\Lambda^+$)
  interference. The errors quoted originate from the limited MC statistics.}
\label{tab:lambda_yields}
\centering
\scriptsize\addtolength{\tabcolsep}{2pt}
% %\small\addtolength{\tabcolsep}{-2pt}
%\resizebox{\columnwidth}{!}{
\begin{tabular}{c r@{\ $\pm$\ }l r@{\ $\pm$\ }l r@{\ $\pm$\ }l r@{\ $\pm$\ }l r@{\ $\pm$\ }l r@{\ $\pm$\ }l r@{\ $\pm$\ }l r@{\ $\pm$\ }l r@{\ $\pm$\ }l r@{\ $\pm$\ }l r@{\ $\pm$\ }l}
\hline
\hline
$m_{\mu\mu}$ [\GeV{}] \T \B & \multicolumn{2}{c}{150-170} & \multicolumn{2}{c}{170-200} & \multicolumn{2}{c}{200-240} & \multicolumn{2}{c}{240-300} & \multicolumn{2}{c}{300-400} & \multicolumn{2}{c}{400-550} & \multicolumn{2}{c}{550-800} & \multicolumn{2}{c}{800-1200} & \multicolumn{2}{c}{1200-2000} \\
\hline
$\Lambda^-$~=~3 \TeV{} \T & 19.1 & 0.5 & 15.7 & 0.4 & 11.2 & 0.4 & 8.5 & 0.3 & 7.9 & 0.3 & 6.0 & 0.3 & 6.5 & 0.3 & 5.1 & 0.2 & 3.0 & 0.2  \\
$\Lambda^-$~=~4 \TeV{} \T & 18.8 & 0.4 & 14.3 & 0.4 & 10.0 & 0.3 & 6.5 & 0.2 & 5.0 & 0.2 & 3.0 & 0.2 & 2.3 & 0.2 & 1.5 & 0.1 & 1.1 & 0.1 \\
$\Lambda^-$~=~5 \TeV{} \T & 17.4 & 0.4 & 14.3 & 0.4 & 9.4 & 0.3 & 6.2 & 0.2 & 4.3 & 0.2 & 2.0 & 0.1 & 1.3 & 0.1 & 0.7 & 0.1 & 0.4 & 0.1 \\
$\Lambda^-$~=~7 \TeV{} \T \B & 17.3 & 0.4 & 13.8 & 0.4 & 9.3 & 0.3 & 6.3 & 0.2 & 3.3 & 0.2 & 1.3 & 0.1 & 0.6 & 0.1 & 0.2 & 0.0 & 0.1 & 0.0 \\
\hline
$\Lambda^+$~=~2 \TeV{} \T & 21.6 & 0.6 & 19.3 & 0.6 & 15.8 & 0.5 & 15.2 & 0.5 & 21.2 & 0.6 & 21.6 & 0.6 & 25.5 & 0.6 & 21.4 & 0.6 & 15.1 & 0.5 \\
$\Lambda^+$~=~3 \TeV{} \T & 18.6 & 0.4 & 15.2 & 0.4 & 10.1 & 0.3 & 7.2 & 0.3 & 5.5 & 0.2 & 4.6 & 0.2 & 5.3 & 0.2 & 4.3 & 0.2 & 3.1 & 0.2 \\
$\Lambda^+$~=~4 \TeV{} \T & 18.2 & 0.4 & 14.3 & 0.4 & 8.8 & 0.3 & 6.1 & 0.2 & 3.6 & 0.2 & 2.1 & 0.1 & 1.6 & 0.1 & 1.5 & 0.1 & 0.8 & 0.1 \\
$\Lambda^+$~=~5 \TeV{} \T \B & 18.5 & 0.4 & 13.6 & 0.3 & 8.8 & 0.3 & 5.4 & 0.2 & 2.9 & 0.2 & 1.6 & 0.1 & 0.9 & 0.1 & 0.5 & 0.1 & 0.3 & 0.1 \\
\hline
\hline
\end{tabular}
%}
\end{table*}

\begin{figure}
  \includegraphics[width=0.48\textwidth]{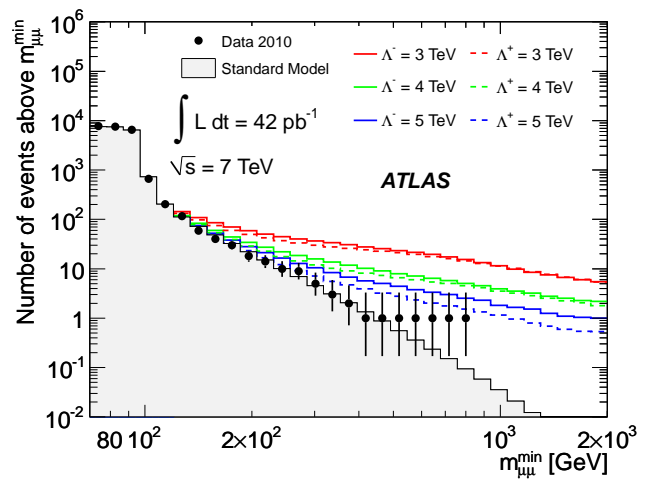}
  \caption{\label{fig:dimuon_mass_integral} Distribution of the number of events
  with dimuon mass above $m_{\mu\mu}^{\min}$ for data (points)
  and Monte Carlo (histograms). The SM prediction is shown as the shaded grey histogram,
  whereas the solid (dashed) histograms correspond to the expected distributions
  in the presence of contact interactions with various scales $\Lambda$ for
  constructive (destructive) interference.}
\end{figure}

% statistical analysis
To estimate the level of agreement between the observed mass spectrum and
the SM prediction, a large ensemble of SM-only pseudo-experiments was generated.
For each such pseudo-experiment, a binned likelihood was computed to quantify
the deviation from the SM expectation.
In 56\% of these pseudo-experiments, the deviation was found to be
more significant than that observed in the data for the signal region, 
indicating good consistency between the data and the predicted spectrum.
This level of agreement is illustrated in 
Fig.~\ref{fig:dimuon_mass_integral}, which shows the number of events above
a minimum mass $m_{\mu\mu}^{\min}$. 
Since no significant deviation is observed in the dimuon mass spectrum, 
we proceed with setting a limit on the energy scale $\Lambda$ using a Bayesian method. 
Here, the prior probability distribution is chosen to be flat in $1/\Lambda^2$,
motivated by the form of Eq.~(\ref{cross_section}).
Systematic uncertainties are incorporated in the limit setting by
treating them as nuisance parameters ($\bar{\nu}$) that are marginalized in the calculation
of the posterior probability ${\cal P}$.
The 95\% confidence level limit is then obtained by finding the value 
$\Lambda_{\rm lim}$ that satisfies
$\int_0^{\theta_{\rm lim}} \text{$\cal P$} (\theta \mid \bar{n},\bar{\nu}) \: d\theta = 0.95$,
where $\theta = 1/\Lambda^2$ and $\bar{n}$ represents the observed number of 
events in the mass bins above 150~\GeV{}, 
with bin boundaries as defined in Table~\ref{tab:dimuon_yields}.
Table~\ref{tab:lambda_yields} shows the expected number of events in each
mass bin within the signal region for different scales $\Lambda$, as used
in the calculation of the posterior probability.

% systematics
Systematic errors are of both theoretical and experimental origins.
Because the expected event yields are normalized to the $Z^0$ peak region,
only momentum- or mass-dependent uncertainties are relevant.
Theoretical uncertainties include PDF variations evaluated using the
{\sc mstw2008} PDF error set~\cite{MSTW} in the absence of a full error set
for the {\sc mrst2007} LO$^\ast$ PDF.
This choice leads to conservative uncertainties in the event yields
that grow from 3\% at the $Z^0$ pole
to 6\% (9\%) at a mass of 1~\TeV{} (1.5~\TeV{}).
A cross-check was made by computing cross sections for both
{\sc mstw2008} and {\sc cteq}~6.6 PDFs for a wide range of dimuon masses.
Differences between the two choices of PDF were always found to be smaller
than the assigned uncertainty obtained from the {\sc mstw2008} PDF set.
The QCD K-factor uncertainty in the DY and DY+CI cross sections is taken to be 
the difference between NNLO and NLO DY cross sections as a function of dimuon mass.
The electroweak K-factor uncertainty in the DY cross section is taken to be 
the entire magnitude of the correction relative to the LO cross section.
Uncertainties in the QCD (electroweak) K-factor are mass dependent; for example,
they amount to 3.0\% (4.5\%) at a mass of 1~\TeV{}.
Uncertainties in the $\ttbar$, diboson and $W$+jets cross sections have
a negligible impact on the limit.
Finally, the statistical error of the DY+CI MC 
(shown in Table~\ref{tab:lambda_yields}) is included as
a source of systematic error.

The MC simulation is used to determine all acceptance and efficiency effects.
Therefore, detailed comparisons between data and Monte Carlo were performed
to make sure that the simulation models the data well for our
choice of muon track selection criteria, especially at higher $p_T$.
Experimental uncertainties arise from the slight $p_T$-dependence of
muon efficiencies and from the impact of the intrinsic detector spatial resolution
on the momentum resolution.
At transverse momenta above 200~\GeV{}, 
radiative losses due to bremsstrahlung in the detector material begin
to affect the muon track pattern recognition.
An uncertainty of 3\% per \TeV{} is assigned to the muon efficiency to conservatively account
for the small $p_T$ dependence predicted by the simulation.
Muon momentum resolution at high $p_T$ is most affected by 
the quality of the muon spectrometer alignment. 
The latter has been studied with high-momentum cosmic ray muons
traversing the center of the detector. It has also been studied in collision data
with muons passing through detector regions with overlapping muon spectrometer chambers,
thereby providing independent track fits from the redundant sets of hits in
neighboring chambers and allowing the impact of the alignment of adjacent detector
regions to be measured.
Curvature smearing parameters derived from these studies are found to be
$\delta(q/p_T) = 0.18 \pm 0.04$ \TeV{}$^{-1}$ for $|\eta| < 2.0$ and
$\delta(q/p_T) = 0.7 \pm 0.2$ \TeV{}$^{-1}$ for $|\eta| > 2.0$,
where $q$ is the charge of the muon track.
These parameters reflect the current level of understanding of
the detector alignment and are expected to decrease with further data taking.
We take the full magnitude of these smearing corrections as the systematic
uncertainty in the momentum resolution.

Using the Bayesian method described above, 
the expected 95\% C.L. lower limits on the scale $\Lambda$ are found to be
$\explimitminus \pm 0.3$ \TeV{} and $\explimitplus \pm 0.3$ \TeV{} for
constructive and destructive interference, respectively.
The quoted uncertainty range is estimated with a large set of pseudo-experiments
and corresponds to a 68\% range around the median value of all the limits 
obtained from those pseudo-experiments.
Systematic errors are already folded into the limit setting procedure
and result in a decrease of the limit by about 0.1~\TeV{}.
The dominant source of uncertainty originates from the limited signal MC statistics.
For the selected data sample, 
we set the following limits at 95\% C.L.:
$\Lambda^- > \obslimitminus$~\TeV{} for constructive interference and
$\Lambda^+ > \obslimitplus$~\TeV{} for destructive interference in the LLIM.
These values are compatible with the expected limits.

% conclusions
To conclude, a search for contact interactions has been carried out
in a sample of dimuon events recorded by the ATLAS detector
in $pp$ collisions from the LHC at $\sqrt{s} = 7$~\TeV{}.
No significant deviation from the Standard Model is observed in the dimuon mass
spectrum obtained from a data sample corresponding to an integrated luminosity 
of $\integlumi$~pb$^{-1}$.
Limits placed on the energy scale $\Lambda$ are the most stringent to date
for $\mu \mu q q$ contact interactions.

%\section{Acknowledgements}

We thank CERN for the very successful operation of the LHC, as well as the
support staff from our institutions without whom ATLAS could not be
operated efficiently.

We acknowledge the support of ANPCyT, Argentina; YerPhI, Armenia; ARC,
Australia; BMWF, Austria; ANAS, Azerbaijan; SSTC, Belarus; CNPq and FAPESP,
Brazil; NSERC, NRC and CFI, Canada; CERN; CONICYT, Chile; CAS, MOST and
NSFC, China; COLCIENCIAS, Colombia; MSMT CR, MPO CR and VSC CR, Czech
Republic; DNRF, DNSRC and Lundbeck Foundation, Denmark; ARTEMIS, European
Union; IN2P3-CNRS, CEA-DSM/IRFU, France; GNAS, Georgia; BMBF, DFG, HGF, MPG
and AvH Foundation, Germany; GSRT, Greece; ISF, MINERVA, GIF, DIP and
Benoziyo Center, Israel; INFN, Italy; MEXT and JSPS, Japan; CNRST, Morocco;
FOM and NWO, Netherlands; RCN, Norway; MNiSW, Poland; GRICES and FCT,
Portugal; MERYS (MECTS), Romania; MES of Russia and ROSATOM, Russian
Federation; JINR; MSTD, Serbia; MSSR, Slovakia; ARRS and MVZT, Slovenia;
DST/NRF, South Africa; MICINN, Spain; SRC and Wallenberg Foundation,
Sweden; SER, SNSF and Cantons of Bern and Geneva, Switzerland; NSC, Taiwan;
TAEK, Turkey; STFC, the Royal Society and Leverhulme Trust, United Kingdom;
DOE and NSF, United States of America.

The crucial computing support from all WLCG partners is acknowledged
gratefully, in particular from CERN and the ATLAS Tier-1 facilities at
TRIUMF (Canada), NDGF (Denmark, Norway, Sweden), CC-IN2P3 (France),
KIT/GridKA (Germany), INFN-CNAF (Italy), NL-T1 (Netherlands), PIC (Spain),
ASGC (Taiwan), RAL (UK) and BNL (USA) and in the Tier-2 facilities
worldwide.

%%%%%%%%%%%%%%%%%%%%%%%%%%%%%%%%%%%%%%%%%%%%%%%%%%%%%%%%%%%%%%%%%%%%%%%%%%%%%%%
% Bibliography
%%%%%%%%%%%%%%%%%%%%%%%%%%%%%%%%%%%%%%%%%%%%%%%%%%%%%%%%%%%%%%%%%%%%%%%%%%%%%%%

\clearpage

\begin{widetext}

% ATLAS Collaboration author list for 21-MAR-2011
% Data extracted on 05-Apr-2011 for paperid 76
%\documentclass[11pt]{article}
%\usepackage{a4wide}\begin{document}
\begin{flushleft}
{\Large The ATLAS Collaboration}

\bigskip

G.~Aad$^{\rm 48}$,
B.~Abbott$^{\rm 111}$,
J.~Abdallah$^{\rm 11}$,
A.A.~Abdelalim$^{\rm 49}$,
A.~Abdesselam$^{\rm 118}$,
O.~Abdinov$^{\rm 10}$,
B.~Abi$^{\rm 112}$,
M.~Abolins$^{\rm 88}$,
H.~Abramowicz$^{\rm 153}$,
H.~Abreu$^{\rm 115}$
% ...
\bigskip

$^{1}$ University at Albany, Albany NY, United States of America\\
% ...
$^{10}$ Institute of Physics, Azerbaijan Academy of Sciences, Baku, Azerbaijan\\
$^{11}$ Institut de F\'isica d'Altes Energies and Universitat Aut\`onoma  de Barcelona and ICREA, Barcelona, Spain\\
% ...
$^{48}$ Fakult\"{a}t f\"{u}r Mathematik und Physik, Albert-Ludwigs-Universit\"{a}t, Freiburg i.Br., Germany\\
$^{49}$ Section de Physique, Universit\'e de Gen\`eve, Geneva, Switzerland\\
% ...
$^{88}$ Department of Physics and Astronomy, Michigan State University, East Lansing MI, United States of America\\
% ...
$^{111}$ Homer L. Dodge Department of Physics and Astronomy, University of Oklahoma, Norman OK, United States of America\\
$^{112}$ Department of Physics, Oklahoma State University, Stillwater OK, United States of America\\
% ...
$^{115}$ LAL, Univ. Paris-Sud and CNRS/IN2P3, Orsay, France\\
% ...
$^{118}$ Department of Physics, Oxford University, Oxford, United Kingdom\\
% ...
$^{153}$ Raymond and Beverly Sackler School of Physics and Astronomy, Tel Aviv University, Tel Aviv, Israel\\
% ...
$^{s}$ Also at California Institute of Technology, Pasadena CA, United States of America\\
% ...
\end{flushleft}

%\end{document}
\end{widetext}


\begin{thebibliography}{99}
\bibitem{coordinates}
ATLAS uses a right-handed coordinate system with its origin at the nominal
interaction point (IP) in the center of the detector and the $z$-axis along
the beam pipe. The $x$-axis points from the IP to the center of the LHC ring,
and the $y$-axis points upward. Cylindrical coordinates $(r,\phi)$
are used in the transverse plane, $\phi$ being the azimuthal angle around
the beam pipe. The pseudorapidity is defined in terms of the polar
angle $\theta$ as $\eta$~=~$-$~$\ln\tan(\theta /2)$.
\end{thebibliography}
\end{document}